\documentclass[aps,pre,showpacs,noshowkeys,amsmath,amssymb,amsfonts,superscriptaddress,longbibliography,reprint]{revtex4-1}
\usepackage[english]{babel}

\usepackage{graphicx}
\usepackage{bm}
\usepackage{physics}
\usepackage{mathtools}
\usepackage{gensymb}

\setcitestyle{super}
\usepackage{caption}
\usepackage{subcaption}
\DeclareCaptionLabelSeparator{bar}{~\rule[-0.4ex]{0.2ex}{1em}~}
\DeclareCaptionLabelFormat{subfor}{\textbf{#2}}
\usepackage{xcolor}
\definecolor{UBcolor}{HTML}{007CC1}
\usepackage[colorlinks=true,pdfnewwindow=true,linkcolor=UBcolor,citecolor=UBcolor,urlcolor=UBcolor,breaklinks=true,linktocpage]{hyperref}
\usepackage[all]{hypcap}
\usepackage[nameinlink,capitalise]{cleveref}
\crefname{SI section}{SI Section}{SI Sections}
\Crefname{SI section}{SI Section}{SI Sections}
\begin{document}

\title{Life at low Reynolds number isn't such a drag}

\author{Sujit S. Datta}
\email{ssdatta@caltech.edu}
\affiliation{Division of Chemistry and Chemical Engineering, California Institute of Technology, Pasadena, CA 91125, USA}

\date{\today}

\begin{abstract}
    \noindent The following is an unedited version of two short articles that are forthcoming in \emph{Nature Chemical Engineering}. Inspired by Purcell's classic lecture ``Life at low Reynolds number'', I discuss how scaling arguments, dimensional analysis, and fundamental concepts from chemical engineering science can be used to quantitatively describe microbial swimming---thereby helping to better understand biological systems and inspiring new engineering advances in turn.
\end{abstract}

\maketitle

\noindent \textbf{(I) Individuals.}
\noindent Imagine swimming in a pool on a hot summer's day. What distance
\(\delta_{\mathrm{macro}}\) do you coast after each stroke that propels you at a
speed \(V\)? The drag force \(f_{\mathrm{drag}}\sim\rho V^{2} L^{2}\)
exerted by the surrounding fluid causes you to decelerate by an amount
\(\sim V^{2}/\delta_{\mathrm{macro}}\); here, \(L\) represents your overall size
and \(\rho\sim m/L^{3}\) is your mass (\(m\)) density. Thus, Newton's
second law dictates that \(f_{\mathrm{drag}}\sim mV^{2}/\delta_{\mathrm{macro}}\), or,
\(\delta_{\mathrm{macro}}\sim L\). That is, you coast a distance comparable to
your size, on the order of a meter.

How about for swimming microbes, which are much smaller and slower than
us? In this case, fluid dynamics tells us that \(f_{\mathrm{drag}}\sim\eta VL\),
where \(\eta\) is the fluid dynamic viscosity. Newton's second law then
says that \(\eta VL\sim mV^{2}/\delta_{\mathrm{micro}}\), or,
\(\delta_{\mathrm{micro}}\ \sim\ mV/(\eta L)\). This coasting distance can be
much smaller than the size of the microbe itself: for the
commonly-studied bacterium \emph{E. coli} in water, $m\simeq1$~pg, $V\simeq10$~$\mu$m/s, $\eta\simeq1$~mPa-s, $L\simeq1$~$\mu$m, yielding a
coasting distance of 0.1 Angstroms --- a tenth the size of a hydrogen
atom! The viscous drag of the surrounding fluid plays such a dominant
role for microbes that it stops them from coasting essentially
instantaneously. An interesting consequence of this fact is that
microbes cannot use reciprocal motion---i.e., motion that looks the same
when played in reverse, as in the case of a scallop repeatedly opening
and closing its shell\footnote{The scallop is just an illustrative example; in reality, it likely closes its shell sufficiently fast so that $\mathrm{Re}>1$.}---to translate large distances. They would just
keep going back and forth and wouldn't get anywhere! Researchers are
working to engineer systems that break this so-called ``scallop
theorem'',\footnote{E. Lauga, ``Life around the scallop theorem'',
  \emph{Soft Matter}, 7, 3060 (2011).} potentially helping to establish
new classes of self-propelling micromachines.

Taking the ratio between the macroscopic and microscopic coasting
distances highlights this increasing relative importance of viscous drag
at small length scales, yielding a dimensionless parameter that chemical
engineers know well:~the Reynolds number
\({\mathrm{Re} = \delta}_{\mathrm{micro}}/\delta_{\mathrm{macro}} = mV/\left( {\eta L}^{2} \right) = \rho VL/\eta\).
As illustrated by the simple calculations above, life at low Reynolds
number (\(\mathrm{Re} < 1\)) is very different from the world of large Reynolds
number (\(\mathrm{Re} > 1\)) that we humans typically inhabit.

These calculations were beautifully described by Edward Purcell, winner
of the Nobel Prize in Physics for discovering nuclear magnetic
resonance, in a 1976 lecture to a distinguished group of physicists
titled ``Life at low Reynolds number''.\footnote{E. M. Purcell, ``Life at
  low Reynolds number'', \emph{Am. J. Phys.}, 45, 3 (1977).} While his lecture was important in establishing an
intellectual framework for thinking about the physics of swimming
microbes, it could have arguably been a chemical engineering lecture instead: after all, it uses scaling arguments and ideas from dimensional
analysis to compare key length, time, and force scales---just as good
chemical engineers do. Indeed, as I highlight below, the contemporary
research areas that have emanated from Purcell's original lecture apply
fundamental concepts from chemical engineering science---mass and energy
balances, thermodynamics, transport phenomena\footnote{J. Happel and H. Brenner, \emph{Low Reynolds number hydrodynamics.} Prentice Hall, Englewood Cliffs, NJ (1967).}, and chemical
kinetics---to shed new light on problems at the interface of
engineering, biology, chemistry, and physics, and in turn, seek
inspiration from these fields to engineer new forms of synthetic
microswimmers for e.g., drug delivery and environmental sensing. They
are therefore of direct interest to modern-day researchers in chemical
engineering and represent an important frontier for the future of the field.

For example: What if the microbe in our thought experiment was swimming
in a polymeric fluid instead of just water? This is a question of clear
biological importance: microbes often inhabit, and cause infections in,
polymeric fluids such as mucus in the lungs and cervicovaginal tract. In
this case, the swimming microbe generates an elastic stress
\(\sim\ \eta\lambda(V/L)^{2}\) in addition to the viscous stress
\(\sim\ \eta V/L\); here, \(\lambda\) is the time scale over which
elastic stresses relax in such viscoelastic fluids and \(V/L\)
characterizes the fluid strain rate generated by swimming. Comparing
these stresses yields a new dimensionless parameter: the Weissenberg
number \(\mathrm{Wi} = \lambda V/L\).\footnote{R. J. Poole, ``The Deborah and
  Weissenberg numbers'', The British Society of Rheology, \emph{Rheology
  Bulletin}, 53, 32 (2012).} When \(\mathrm{Wi} > 1\), swimming is fast enough to
stretch the flexible polymers in the fluid. The resistance generated by
polymer elasticity feeds back on the microbe and can modify how it swims
in unexpected ways---for example, causing it to wobble less and speed
up.\footnote{A. Martínez-Calvo, C. Trenado-Yuste, S. S. Datta, ``Active
  transport in complex environments'', in \emph{Out-of-Equilibrium Soft
  Matter: Active Fluids}, RSC press (2023).} Continuing to explore the
coupling between fluid rheology and micro-scale swimming, both for
biological microbes as well as engineered microswimmers, is a vibrant
area of current research at the interface of fluid dynamics, biology,
and the physics of actively moving systems (``active matter'').

One may ask: Why do microbes swim in the first place? The answer may
seem obvious: It helps them take up food molecules from their
surroundings, like little Pac-Men! But ideas from chemical engineering
can help examine the validity of this assumption more quantitatively, as Berg \& Purcell
did in a follow-up 1977 paper.\footnote{H. C. Berg and E. M. Purcell,
  ``Physics of chemoreception'', \emph{Biophysical journal,} 20, 193,
  (1977).} As a microbe swims through fluid, it drags the fluid within a
distance \(\sim d\) from its surface along with it --- reflecting the
no-slip condition of fluids at solid surfaces. So, surrounding nutrient
molecules get pushed aside as well, and can only reach the surface of
the cell, where they can be taken up, via thermal diffusion with a
diffusivity \(D\). This process takes a considerable amount of time
\({\sim d}^{2}/D\) compared to that of simple advective transport,
\(\sim d/V\). Comparing these time scales yields a new dimensionless
parameter:~the Péclet number \(\mathrm{Pe} = dV/D\). Berg \& Purcell's numerical
calculations of the flow field around an idealized spherical microbe,
which built on earlier calculations by chemical
engineers,\footnote{S. K. Friedlander, ``Mass and heat transfer to single
  spheres and cylinders at low Reynolds numbers'', \emph{AIChE journal}
  3, 43 (1957).}\textsuperscript{,}\footnote{A. Acrivos and T. D.
  Taylor, ``Heat and mass transfer from single spheres in Stokes flow'',
  \emph{Physics of Fluids} 5, 387 (1962).} showed that the thickness \(d\)
of the region surrounding the cell that has \(\mathrm{Pe} < 1\), through which
nutrient transport is diffusion-limited, does not shrink much as
swimming speed increases. Thus, they concluded that ``in a uniform
medium motility cannot significantly increase the cell's acquisition of
material'' ---~suggesting that microbial swimming has a more complex
biological function.\footnote{A caveat is that, when nutrients are non-uniformly distributed, cells can use chemotaxis to swim up gradients to find more nutrients.} Ongoing research is building on these results to
consider more complex microbial shapes and swimming behaviors, not just
in the context of nutrient acquisition, but also interactions with other
suspended materials such as antimicrobials.\footnote{C. Lohrmann, C. Holm,
  S. S. Datta, ``Influence of bacterial swimming and hydrodynamics on
  infection by phages'', \emph{Soft Matter} 20, 4785 (2024).}\\

\noindent \textbf{(II) Collectives.} Our discussion thus far has focused on individual microswimmers.
However, in nature and many engineering applications, microbes (and
other microswimmers) move in collectives in which they alter each
other's motion via the fluid they push aside, as well as by simply
bumping into each other. What are the emergent consequences of these
interactions, and how can they be quantitatively described? There is
tremendous current interest in addressing this question,\footnote{M. C.
  Marchetti, J. F. Joanny, S. Ramaswamy, T. B. Liverpool, J. Prost, M.
  Rao, R. A. Simha, ``Hydrodynamics of soft active matter'',
  \emph{Reviews of modern physics}, 85, 1143 (2013).} although its roots
date back to a classic 1969 paper by the renowned chemical engineers
Bruce Finlayson and L. E. ``Skip'' Scriven in which they explored the
motion of fluids due to ``active stresses'' generated by, or in, living
organisms such as slime molds.\footnote{B. A. Finlayson, and L. E.
  Scriven, ``Convective instability by active stress'', \emph{Proceedings
  of the Royal Society of London. A. Mathematical and Physical Sciences}
  310, 183 (1969).}

The microbe from our thought experiment, for example, generates a force
equal and opposite to \(f_{\mathrm{drag}}\sim\eta VL\) to swim (so-called
``force-free'' motion at low Reynolds number). Say it swims for a
duration \(\tau_{R}\) (typically $\sim10$~s) before changing
direction~either by choice or inevitable Brownian reorientation. The
corresponding amount of work done is then \(\sim\eta V^{2}L\tau_{R}\)
---approximately 1000 times larger than thermal energy, \(k_{B}T\), at
room temperature! Hence, for a collective of microswimmers at a dilute
number density \(n\), by analogy to the Brownian osmotic pressure
\(\Pi\sim nk_{B}T\), one can define an additional ``swim pressure''
\(\Pi^{\mathrm{swim}}\sim n \cdot \eta V^{2}L\tau_{R}\). (It is tempting to
define an ``effective temperature''
\(T_{\mathrm{eff}}\ \sim\eta V^{2}L\tau_{R}/k_{B}\) for this system, but it
remains unclear if and when such a description is appropriate in
actively moving collectives.) Just as in the kinetic theory of gases,
molecular collisions on surrounding boundaries generate our familiar
notion of pressure, the collisions between microswimmers and boundaries
generates this new kind of active pressure.\footnote{S. C. Takatori and
  J. F. Brady, ``Forces, stresses and the (thermo?) dynamics of active
  matter'', \emph{Current Opinion in Colloid \& Interface Science}, 21,
  24 (2016).}

Fascinating new effects can emerge in even denser collectives of active
microswimmers. For example, when swimming is highly persistent ---~i.e.,
the duration \(\tau_{R}\) is long compared to the time needed to swim a
body length, \(L/V\), or equivalently, the reorientation Péclet number
\({\mathrm{Pe}}_{R} = L/(V\tau_{R})\) is small ---~the microswimmers impede each
other's motion, causing a collective to undergo ``motility-induced phase
separation'' (MIPS) into dense and dilute phases. We learn in
undergraduate thermodynamics that phase separation in ``passive''
equilibrium systems typically requires attractive interactions between
the constituents; in stark contrast, this active form of phase
separation arises from purely repulsive interactions. And remarkably,
despite being out of thermodynamic equilibrium, MIPS can be
mathematically described using models inspired by the phase separation
of thermally equilibrated passive constituents.\footnote{M. E. Cates \&
  J. Tailleur, ``Motility-induced phase separation'', \emph{Annu. Rev.
  Condens. Matter Phys.}, 6, 219 (2015).} Further exploring how ideas
from equilibrium thermodynamics and statistical mechanics can---or
cannot---be used to describe active collectives, as well as using these
insights to develop ways to e.g., control their spatial organization and
put them to work for new forms of actuation, continues to be an exciting
frontier of research.

We also learn as undergraduates that fluid flows at low Reynolds number
are laminar. Despite living at low Reynolds number,
dense collectives of actively moving microswimmers can overturn this
paradigm. A prominent example is the phenomenon of ``active
turbulence''. Because many microswimmers (like bacteria, sperm, and even
motor-driven biopolymer assemblies) are elongated in shape, they tend to
align when close enough to each other---just like the molecules in a
nematic liquid crystal.\footnote{D. L. Koch \& G. Subramanian,
  ``Collective hydrodynamics of swimming microorganisms: living
  fluids'', \emph{Annual Review of Fluid Mechanics}, 43, 637 (2011).}
Deviations from this alignment between neighbors are penalized, as
described by a Frank elastic constant \(K\). As a result, microswimmers
in a dense collective often move in coherent swirls that, as
they interact, give rise to large-scale chaotic flows reminiscent of
turbulence! Ongoing research continues to explore the similarities and
differences between this intriguing phenomenon and typical
inertial turbulence. For example, just as the transition from
laminar to inertial turbulence can be parameterized using the Reynolds
number, recent work hints that a distinct dimensionless parameter
---~the ``activity number'' \(A\) ---~may similarly described the
transition to active turbulence.\footnote{R. Alert, J. Casademunt, J. F.
  Joanny, ``Active turbulence'', \emph{Annual Review of Condensed Matter
  Physics}, 13, 143 (2022).} One way to define this parameter is as
\(A = L_{s}/\sqrt{K/\zeta}\), which compares the system size \(L_{s}\)
to the length scale \(\sqrt{K/\zeta}\ \)above which active swimming
overcomes the restoring force of nematic elasticity; here, \(\zeta\) is
an active stress coefficient, which can be thought of as being
\(\sim\Pi^{\mathrm{swim}}\), for example.\footnote{A related parameter is the Ericksen number, $\mathrm{Er}=\eta V L/K$, which compares the force generated by swimming $\eta V L$ to the elastic restoring force $K$.} In addition to unraveling these
fundamental puzzles, an important direction for chemical engineering
research is to develop ways to control active turbulence and harness it
for e.g., improved solute mixing.

  In closing, I note that our discussion thus far has focused on the
  fascinating behaviors of microbes swimming in bulk \emph{fluids} at low
  Reynolds number. But in many cases, microbes instead swim through
  elastic \emph{solids}, such as the extracellular polymer matrix in a
  bacterial biofilm. In this case, confinement by the elastic matrix
  changes how microbes swim.\footnote{T. Bhattacharjee \& S. S. Datta,
    ``Bacterial hopping and trapping in porous media'', \emph{Nature
    Communications} 10, 2075 (2019).} This change can be described by two new dimensionless parameters: the Deborah number $\mathrm{De}=t_{\mathrm{mat}}/t_{\mathrm{swim}}$ comparing the relaxation time scale of the matrix $t_{\mathrm{mat}}$ to a characteristic time scale of swimming $t_{\mathrm{swim}}$, and the Benes number $\mathrm{Bn}=L_{\mathrm{mat}}/L$ comparing the confinement length scale imposed by the matrix $L_{\mathrm{mat}}$ to the size of the microbe $L$.\footnote{S. E. Spagnolie and P. T. Underhill, ``Swimming in complex fluids'', \emph{Annual Review of Condensed Matter Physics}, 14, 381 (2023).} But also, the active stress generated by
  swimming can deform the matrix itself---unexpectedly leading to
  formation of self-sustained elastic standing waves.\footnote{H. Xu, Y.
    Huang, R. Zhang, Y. Wu, ``Autonomous waves and global motion modes
    in living active solids'', \emph{Nature Physics}, 19, 46 (2023).}
  One might imagine that the dimensionless parameter \(\Pi^{\mathrm{swim}}/G\),
  where \(G\) is the shear modulus of the elastic matrix, characterizes
  the onset of this behavior, although such a parameter has not, to my
  knowledge, yet been identified. Investigating the unusual properties
  of such ``active solids'' continues to be a fascinating direction for
  ongoing research.

The eminent chemical engineer Rutherford ``Gus'' Aris wrote in
1973:\footnote{R. Aris, ``Some Interactions between Problems in Chemical
  Engineering and the Biological Sciences'', in \emph{Environmental
  Engineering: A Chemical Engineering Discipline}, pp. 215-225,
  Dordrecht: Springer Netherlands (1973).} ``\ldots let us get
acquainted with our colleagues working in physiology, biology, zoology,
genetics, microbiology and like disciplines for though their problems
are infinitely more complex than engineering ones they offer a
challenging and important field for our efforts.'' Over a half century
later, these words ring truer than ever. As the discussion above
hopefully highlights, the tools and perspective of chemical engineering
have immense potential to help shed new light on biological systems,
such as swimming microbes; and these systems in turn present
opportunities for new fundamental discoveries and technological advances
in chemical engineering. Life at low Reynolds number, it turns out,
isn't such a drag.


\begin{thebibliography}{21}%
\makeatletter
\providecommand \@ifxundefined [1]{%
 \@ifx{#1\undefined}
}%
\providecommand \@ifnum [1]{%
 \ifnum #1\expandafter \@firstoftwo
 \else \expandafter \@secondoftwo
 \fi
}%
\providecommand \@ifx [1]{%
 \ifx #1\expandafter \@firstoftwo
 \else \expandafter \@secondoftwo
 \fi
}%
\providecommand \natexlab [1]{#1}%
\providecommand \enquote  [1]{``#1''}%
\providecommand \bibnamefont  [1]{#1}%
\providecommand \bibfnamefont [1]{#1}%
\providecommand \citenamefont [1]{#1}%
\providecommand \href@noop [0]{\@secondoftwo}%
\providecommand \href [0]{\begingroup \@sanitize@url \@href}%
\providecommand \@href[1]{\@@startlink{#1}\@@href}%
\providecommand \@@href[1]{\endgroup#1\@@endlink}%
\providecommand \@sanitize@url [0]{\catcode `\\12\catcode `\$12\catcode `\&12\catcode `\#12\catcode `\^12\catcode `\_12\catcode `\%12\relax}%
\providecommand \@@startlink[1]{}%
\providecommand \@@endlink[0]{}%
\providecommand \url  [0]{\begingroup\@sanitize@url \@url }%
\providecommand \@url [1]{\endgroup\@href {#1}{\urlprefix }}%
\providecommand \urlprefix  [0]{URL }%
\providecommand \Eprint [0]{\href }%
\providecommand \doibase [0]{http://dx.doi.org/}%
\providecommand \selectlanguage [0]{\@gobble}%
\providecommand \bibinfo  [0]{\@secondoftwo}%
\providecommand \bibfield  [0]{\@secondoftwo}%
\providecommand \translation [1]{[#1]}%
\providecommand \BibitemOpen [0]{}%
\providecommand \bibitemStop [0]{}%
\providecommand \bibitemNoStop [0]{.\EOS\space}%
\providecommand \EOS [0]{\spacefactor3000\relax}%
\providecommand \BibitemShut  [1]{\csname bibitem#1\endcsname}%
\let\auto@bib@innerbib\@empty
\bibitem [{Note1()}]{Note1}%
  \BibitemOpen
  \bibinfo {note} {The scallop is just an illustrative example; in reality, it likely closes its shell sufficiently fast so that $\protect \mathrm {Re}>1$.}\BibitemShut {Stop}%
\bibitem [{Note2()}]{Note2}%
  \BibitemOpen
  \bibinfo {note} {E. Lauga, ``Life around the scallop theorem'', \protect \emph {Soft Matter}, 7, 3060 (2011).}\BibitemShut {Stop}%
\bibitem [{Note3()}]{Note3}%
  \BibitemOpen
  \bibinfo {note} {E. M. Purcell, ``Life at low Reynolds number'', \protect \emph {Am. J. Phys.}, 45, 3 (1977).}\BibitemShut {Stop}%
\bibitem [{Note4()}]{Note4}%
  \BibitemOpen
  \bibinfo {note} {J. Happel and H. Brenner, \protect \emph {Low Reynolds number hydrodynamics.} Prentice Hall, Englewood Cliffs, NJ (1967).}\BibitemShut {Stop}%
\bibitem [{Note5()}]{Note5}%
  \BibitemOpen
  \bibinfo {note} {R. J. Poole, ``The Deborah and Weissenberg numbers'', The British Society of Rheology, \protect \emph {Rheology Bulletin}, 53, 32 (2012).}\BibitemShut {Stop}%
\bibitem [{Note6()}]{Note6}%
  \BibitemOpen
  \bibinfo {note} {A. Martínez-Calvo, C. Trenado-Yuste, S. S. Datta, ``Active transport in complex environments'', in \protect \emph {Out-of-Equilibrium Soft Matter: Active Fluids}, RSC press (2023).}\BibitemShut {Stop}%
\bibitem [{Note7()}]{Note7}%
  \BibitemOpen
  \bibinfo {note} {H. C. Berg and E. M. Purcell, ``Physics of chemoreception'', \protect \emph {Biophysical journal,} 20, 193, (1977).}\BibitemShut {Stop}%
\bibitem [{Note8()}]{Note8}%
  \BibitemOpen
  \bibinfo {note} {S. K. Friedlander, ``Mass and heat transfer to single spheres and cylinders at low Reynolds numbers'', \protect \emph {AIChE journal} 3, 43 (1957).}\BibitemShut {Stop}%
\bibitem [{Note9()}]{Note9}%
  \BibitemOpen
  \bibinfo {note} {A. Acrivos and T. D. Taylor, ``Heat and mass transfer from single spheres in Stokes flow'', \protect \emph {Physics of Fluids} 5, 387 (1962).}\BibitemShut {Stop}%
\bibitem [{Note10()}]{Note10}%
  \BibitemOpen
  \bibinfo {note} {A caveat is that, when nutrients are non-uniformly distributed, cells can use chemotaxis to swim up gradients to find more nutrients.}\BibitemShut {Stop}%
\bibitem [{Note11()}]{Note11}%
  \BibitemOpen
  \bibinfo {note} {C. Lohrmann, C. Holm, S. S. Datta, ``Influence of bacterial swimming and hydrodynamics on infection by phages'', \protect \emph {Soft Matter} 20, 4785 (2024).}\BibitemShut {Stop}%
\bibitem [{Note12()}]{Note12}%
  \BibitemOpen
  \bibinfo {note} {M. C. Marchetti, J. F. Joanny, S. Ramaswamy, T. B. Liverpool, J. Prost, M. Rao, R. A. Simha, ``Hydrodynamics of soft active matter'', \protect \emph {Reviews of modern physics}, 85, 1143 (2013).}\BibitemShut {Stop}%
\bibitem [{Note13()}]{Note13}%
  \BibitemOpen
  \bibinfo {note} {B. A. Finlayson, and L. E. Scriven, ``Convective instability by active stress'', \protect \emph {Proceedings of the Royal Society of London. A. Mathematical and Physical Sciences} 310, 183 (1969).}\BibitemShut {Stop}%
\bibitem [{Note14()}]{Note14}%
  \BibitemOpen
  \bibinfo {note} {S. C. Takatori and J. F. Brady, ``Forces, stresses and the (thermo?) dynamics of active matter'', \protect \emph {Current Opinion in Colloid \& Interface Science}, 21, 24 (2016).}\BibitemShut {Stop}%
\bibitem [{Note15()}]{Note15}%
  \BibitemOpen
  \bibinfo {note} {M. E. Cates \& J. Tailleur, ``Motility-induced phase separation'', \protect \emph {Annu. Rev. Condens. Matter Phys.}, 6, 219 (2015).}\BibitemShut {Stop}%
\bibitem [{Note16()}]{Note16}%
  \BibitemOpen
  \bibinfo {note} {D. L. Koch \& G. Subramanian, ``Collective hydrodynamics of swimming microorganisms: living fluids'', \protect \emph {Annual Review of Fluid Mechanics}, 43, 637 (2011).}\BibitemShut {Stop}%
\bibitem [{Note17()}]{Note17}%
  \BibitemOpen
  \bibinfo {note} {R. Alert, J. Casademunt, J. F. Joanny, ``Active turbulence'', \protect \emph {Annual Review of Condensed Matter Physics}, 13, 143 (2022).}\BibitemShut {Stop}%
  \bibitem [{Note18()}]{Note18}%
  \BibitemOpen
  \bibinfo {note} {A related parameter is the Ericksen number, $\mathrm{Er}=\eta V L/K$, which compares the force generated by swimming $\eta V L$ to the elastic restoring force $K$.}\BibitemShut {Stop}%
\bibitem [{Note19()}]{Note19}%
  \BibitemOpen
  \bibinfo {note} {T. Bhattacharjee \& S. S. Datta, ``Bacterial hopping and trapping in porous media'', \protect \emph {Nature Communications} 10, 2075 (2019).}\BibitemShut {Stop}%
\bibitem [{Note20()}]{Note20}%
  \BibitemOpen
  \bibinfo {note} {S. E. Spagnolie and P. T. Underhill, ``Swimming in complex fluids'', \protect \emph {Annual Review of Condensed Matter Physics}, 14, 381 (2023).}\BibitemShut {Stop}%
\bibitem [{Note21()}]{Note21}%
  \BibitemOpen
  \bibinfo {note} {H. Xu, Y. Huang, R. Zhang, Y. Wu, ``Autonomous waves and global motion modes in living active solids'', \protect \emph {Nature Physics}, 19, 46 (2023).}\BibitemShut {Stop}%
\bibitem [{Note22()}]{Note22}%
  \BibitemOpen
  \bibinfo {note} {R. Aris, ``Some Interactions between Problems in Chemical Engineering and the Biological Sciences'', in \protect \emph {Environmental Engineering: A Chemical Engineering Discipline}, pp. 215-225, Dordrecht: Springer Netherlands (1973).}\BibitemShut {Stop}%
\end{thebibliography}
\end{document}